\documentclass[prl,twocolumn,floatfix,showpacs]{revtex4}
\usepackage{amsmath,amsfonts,amssymb}
\usepackage[dvips]{graphicx}
\usepackage{ulem}
\usepackage{color}
\newcommand{\nc}{\newcommand}
\nc{\rnc}{\renewcommand}

\nc{\red}{\textcolor{red}}
\nc{\blue}{\textcolor{blue}}

\nc{\nn}{\nonumber}
\nc{\db}{\displaybreak[0]\\}
\nc{\ds}{\displaystyle}

\rnc{\[}{\left[}
\rnc{\]}{\right]}
\rnc{\(}{\left(}
\rnc{\)}{\right)}
\nc{\lt}{\left\{}
\nc{\rt}{\right\}}

\nc{\om}{\omega}
\nc{\sg}{\sigma}
\nc{\g}{\gamma}
\nc{\lam}{\lambda}
\nc{\eps}{\epsilon}
\rnc{\a}{\alpha}
\rnc{\b}{\beta}
\nc{\vp}{\varphi}
\nc{\kp}{\kappa}
\rnc{\th}{\theta}

\nc{\ch}{\cosh}
\nc{\sh}{\sinh}
\nc{\sech}{\text{sech}}
\nc{\bra}{\langle}
\nc{\ket}{\rangle}
\nc{\up}{\uparrow}
\nc{\down}{\downarrow}
\nc{\vup}{|\!\up\,\ket}
\nc{\vdown}{|\!\down\,\ket}
\nc{\vac}{|\,0\,\ket}
\nc{\vvac}{\bra\,0\,|}
\nc\cd\cdots

\rnc{\i}{{\rm i}}
\rnc{\d}{{\rm d}}

\nc\del\partial

\nc\hll{H}
\nc\hhll{\hat{\mathcal{H}}}
\nc\fcl{f}
\nc\fast{f^\ast}
\nc\hp{\hat{\psi}}
\nc\hpd{\hat{\psi}^\dagger}
\nc\ah{A}
\nc\at{\widetilde{A}}

\begin{document}
%%%%%%%%%%%%%%%%%%%%%%%%%%%%%%%%%%%%%%%%%%%%%%%%%%%%%%%%%%%%%%%%%%%%%%%%%%%
\title{
Quantum-classical correspondence via coherent state 
in integrable field theory
%in the nonlinear Schr{\"o}dinger equation
}
%%%%%%%%%%%%%%%%%%%%%%%%%%%%%%%%%%%%%%%%%%%%%%%%%%%%%%%%%%%%%%%%%%%%%%%%%%%
\author{Jun Sato}
%\ead{jsato@jamology.rcast.u-tokyo.ac.jp}
\address{
Research Center for Advanced Science and Technology, University of Tokyo, \it 
4-6-1 Komaba, Meguro-ku, Tokyo 153-8904, Japan
}
\author{Tsukasa Yumibayashi}
%\ead{tsukasa.yumibayashi@otsuma.ac.jp}
\address{
\it 
Department of Social Information Studies, Otsuma Women's University, 
12 Sanban-cho, Chiyoda-ku, Tokyo 102-8357, Japan
}
%%%%%%%%%%%%%%%%%%%%%%%%%%%%%%%%%%%%%%%%%%%%%%%%%%%%%%%%%%%%%%%%%%%%%%%%%%%
\date{\today}
%%%%%%%%%%%%%%%%%%%%%%%%%%%%%%%%%%%%%%%%%%%%%%%%%%%%%%%%%%%%%%%%%%%%%%%%%%%
%%%%%%%%%%%%%%%%%%%%%%%%%%%%%%%%%%%%%%%%%%%%%%%%%%%%%%%%%%%%%%%%%%%%%%%%%%%
%%%%%%%%%%%%%%%%%%%%%%%%%%%%%%%%%%%%%%%%%%%%%%%%%%%%%%%%%%%%%%%%%%%%%%%%%%%
\begin{abstract}
We consider the problem of quantum-classical correspondence 
in integrable field theories. 
We propose a method to construct a field theoretical coherent state, 
in which the expectation value of the quantum field operator 
exactly coincides with the classical soliton. 
We also discuss the time evolution of this quantum state 
and the instability due to the nonlinearity. 
\end{abstract}
%%%%%%%%%%%%%%%%%%%%%%%%%%%%%%%%%%%%%%%%%%%%%%%%%%%%%%%%%%%%%%%%%%%%%%%%%%%
%%%%%%%%%%%%%%%%%%%%%%%%%%%%%%%%%%%%%%%%%%%%%%%%%%%%%%%%%%%%%%%%%%%%%%%%%%%
\maketitle
%%%%%%%%%%%%%%%%%%%%%%%%%%%%%%%%%%%%%%%%%%%%%%%%%%%%%%%%%%%%%%%%%%%%%%%%%%%
%%%%%%%%%%%%%%%%%%%%%%%%%%%%%%%%%%%%%%%%%%%%%%%%%%%%%%%%%%%%%%%%%%%%%%%%%%%
%%%%%%%%%%%%%%%%%%%%%%%%%%%%%%%%%%%%%%%%%%%%%%%%%%%%%%%%%%%%%%%%%%%%%%%%%%%
%%%%%%%%%%%%%%%%%%%%%%%%%%%%%%%%%%%%%%%%%%%%%%%%%%%%%%%%%%%%%%%%%%%%%%%%%%%
%%%%%%%%%%%%%%%%%%%%%%%%%%%%%%%%%%%%%%%%%%%%%%%%%%%%%%%%%%%%%%%%%%%%%%%%%%%
%\subsection{Introduction}
{\it Introduction. }
The quantum-classical correspondence has been a fundamental problem 
since the foundation of quantum mechanics \cite{qc_correspondence}. 
The problem of how to deduce the classical mechanics from the quantum theory
has been discussed in various ways, for instance, 
the Ehrenfest's theorem \cite{Ehrenfest}, the WKB analysis \cite{WKB}.
Physical quantities in the classical mechanics appears 
as the expectation values in the quantum mechanics. 
It follows that the essential problem in the quantum-classical correspondence 
is to find a quantum state 
in which the expectation value of the canonical variable of the system 
coincides with the classical counterpart. 

The coherent states are known to be quantum states 
which behave classically \cite{coherent_original}. 
In particular, in the case of the harmonic oscillator, 
a coherent state is a localized wave packet 
which oscillates exactly in the same frequency as the classical particle 
without changing its form \cite{coherent_ho}. 
Concerning the integrable system, 
coherent states play a prominent roll 
with respect to the quantum-classical correspondence. 
In \cite{Ichikawa}, KdV soliton was constructed 
as a coherent state of the unharmonic oscillator. 
Moreover, the field theoretical coherent state was constructed in the sine-Gordon model \cite{Aoyama}. 

In this letter, we consider the quantum-classical correspondence in integrable field theories.
The most simple and fundamental example is 
the one-dimensional Bose gas with contact interactions, 
described by the Hamiltonian
\begin{align}
\hll=\sum_{j=1}^{N}\(-\partial_j^2\)+\sum_{1\leq j<k \leq N}2c\delta(x_j,x_k), 
\label{hll}
\end{align}
where $N$ is the number of particle, $c$ is the coupling constant 
and $\del_j:=\dfrac{\del}{\del x_j}$. 
This is a quantum integrable system and 
exact eigenstates and eigenenergies are obtained via the Bethe ansatz method \cite{Lieb-Liniger}. 
In the quantum field description, 
the time evolution of 
the Boson field operator obeys the quantum nonlinear Schr{\"o}dinger (NLS) equation.
In the classical limit where the quantum field operator 
is replaced by a commutative complex scalar field, 
the classical NLS equation is known to be classically integrable 
and has soliton solutions \cite{ZS}. 

Identifying the quantum state corresponding to the classical soliton  
has been a long standing problem. 
In the attractive case $c<0$, 
the classical NLS equation has the bright soliton solution 
and the corresponding quantum state is constructed 
in terms of the bound states associated with the complex Bethe roots called string \cite{Nohl, WKK}. 
In the repulsive case $c>0$, 
the classical solution is the dark soliton. 
It has been argued that the quantum wave packet constructed from 
the superposition of the hole-type excitations 
via the Bethe ansatz is corresponding to the classical dark soliton \cite{SKKD, SKKD2}. 

In the attractive case, 
the bright soliton is obtained from the $N$-particle bound states 
in the limit $N\to\infty$ and $c\to0$ 
while keeping the product $Nc$ finite. 
This can be regarded as a large quantum-number limit. 
Moreover, the time evolution of the quantum state does not obey the law of quantum mechanics. 
The time dependent quantum state is obtained from the Galilean transformation. 
In the repulsive case, the quantum wave packet 
corresponding to the classical dark soliton 
collapses due to the interference of the different energy eigenstates. 

In this letter, we construct a quantum soliton using the coherent state. 
Then we define the quantum and classical time evolutions of this state. 
Comparing their difference, we examine the stability of the quantum soliton state. 
It is known that the nonlinearity violates the stability of coherent states 
\cite{coherent_stable_conditions}. 

%%%%%%%%%%%%%%%%%%%%%%%%%%%%%%%%%%%%%%%%%%%%%%%%%%%%%%%%%%%%%%%%%%%%%%%%%%%
{\it Quantum field theory. }
%\subsection{Quantum field theory.}
The 1D Bose gas \eqref{hll} can be described by 
the Bose field operators satisfying the canonical equal-time commutation relations
\begin{align}
&[\hat{ \psi}(x,t), \hat{ \psi}^{\dagger}(y,t)]=\delta(x-y), \\
&[\hat{ \psi}(x,t), \hat{ \psi}(y,t)]=[\hat{ \psi}^{\dagger}(x,t), \hat{ \psi}^{\dagger}(y,t)]=0. 
\end{align}
The vacuum $\vac$ satisfies
\begin{align}
\hat{ \psi}(x,t)\vac=0,
\quad
\vvac\hat{ \psi}^{\dagger}(x,t)=0,
\qquad
\bra  0 | 0 \ket=1. 
\end{align}
The state space is generated by the successive actions of the creation operator 
$\hpd(x)$ on the vacuum as 
\begin{align}
|\vp_N\ket=\int\d x_1\cdots\d x_N &
\vp_N(x_1,\cdots,x_N)
\nn\\&\times
\hpd(x_1)\cdots\hpd(x_N)\vac, 
\end{align}
where $\vp_N(x_1,\cdots,x_N)$ is the corresponding $N$-body wave function. 
The Hamiltonian $\hhll$
is written in terms of the field operators as
\begin{align}
\hhll=
\int\d x
\[
-\hat{\psi}^\dagger \del_x^2\hat{\psi} 
+
c
\hat{\psi}^\dagger
\hat{\psi}^\dagger
\hat{\psi}
\hat{\psi}
\]. 
\label{hhll}
\end{align}
In the Heisenberg picture, 
the time evolution of the field operator $\hat{\psi}(x,t)$ is given by 
\begin{align}
\i\del_t\hat{\psi}
=[\hat{\psi}, \hhll]
=
-\del_x^2 \hat{\psi} 
+ 2c \hat{ \psi}^{\dagger} \hat{ \psi} \hat{ \psi}, 
\end{align}
which we call the quantum nonlinear Schr{\"o}dinger (NLS) equation. 
The formal solution is explicitly written as
\begin{align}
\hat{\psi}(x,t)
=
e^{\i\hhll t} \hat{\psi}(x) e^{-\i\hhll t}. 
\end{align}

%%%%%%%%%%%%%%%%%%%%%%%%%%%%%%%%%%%%%%%%%%%%%%%%%%%%%%%%%%%%%%%%%%%%%%%%%%%
{\it Classical field theory. }
%\subsection{Classical integrable system}
Replacing the quantum field operators $\hp(x,t)$ and $\hpd(x,t)$ 
by the commutative complex scalar field $\fcl(x,t)$ and $\fast(x,t)$, 
which we call ``{\it classicalization}, 
we obtain the classical NLS equation
\begin{align}
\i\del_t\fcl
=
-\del_x^2 \fcl 
+ 2c \fast \fcl \fcl. 
\label{cNLS}
\end{align}
This can be solved via the inverse scattering method and has a soliton solutions \cite{ZS}. 
The energy functional is obtained through the classicalization in the Hamiltonian \eqref{hhll} as
\begin{align}
E[\fcl,\fast]=\int\d x\[-\fast \del_x^2 \fcl+c\fast\fast\fcl\fcl\], 
\end{align}
in terms of which 
the classical NLS equation \eqref{cNLS} can be recast into the form
\begin{align}
\del_t\fcl=\{\fcl, E\}, \quad
\del_t\fast=\{\fast, E\}, 
\label{cNLS2}
\end{align}
where the Poisson bracket for two functionals is defined by 
\begin{align}
\{F, G\}:=\frac1\i\int\d x
\(
\frac{\delta F}{\delta \fcl}\frac{\delta G}{\delta \fast}
-
\frac{\delta G}{\delta \fcl}\frac{\delta F}{\delta \fast}
\). 
\end{align}
Then it follows that the equal-time canonical relation 
\begin{align}
\{\fcl(x,t), \fast(y,t)\}=\frac1\i\delta(x-y) 
\end{align}
and the time evolution of the physical quantity $F$
\begin{align}
\del_t F=\{F, E\}. 
\end{align}

%%%%%%%%%%%%%%%%%%%%%%%%%%%%%%%%%%%%%%%%%%%%%%%%%%%%%%%%%%%%%%%%%%%%%%%%%%%
{\it Coherent state. }
%\subsection{Coherent state}
Let $\fcl(x, t)$ be the exact soliton solution of the classical  NLS equation 
\eqref{cNLS} or \eqref{cNLS2}. 
We construct a quantum state corresponding to the classical soliton at the initial time $t=0$. 
The main object of this letter is the 
coherent state in the quantum field theory defined as \cite{Aoyama}
\begin{align}
|\fcl \ket:=e^{\ah}\vac, 
\quad
\ah:=\int\d x \[\fcl(x) \hpd(x)-\fast(x) \hp(x)\], 
\label{fket}
\end{align}
where $f(x):=f(x,t=0)$. 
The normalization $\bra\fcl|\fcl\ket=1$ follows from $\ah^\dagger=-\ah$. 
One can easily see that 
\begin{align}
&
[\hp(x), \ah]=\fcl(x), \quad
[\hpd(x), \ah]=\fast(x), 
\\&
\{\fcl(x), \ah\}=\i\hp(x), \quad
\{\fast(x), \ah\}=\i\hpd(x). 
\end{align}
It follows that the coherent state is an eigenvector of $\hat{\psi}(x)$ 
with the eigenvalue $\fcl(x)$
\begin{align}
\hat{\psi}(x)|\fcl \ket=\fcl(x)|\fcl \ket. 
\end{align}
Moreover, as for the expectation values, we have the following relations 
\begin{align}
\bra\fcl|\hp(x)|\fcl\ket=\fcl(x), \quad
\bra\fcl|\hpd(x)|\fcl\ket=\fast(x), 
\end{align}
which means that the coherent state classicalize the field operators $\hp, \hpd$ to the 
scalar fileds $\fcl, \fast$, respectively. 
%The coherent state $|\fcl\ket$ is expanded as
%\begin{align}
%|f \ket
%=
%A\sum_{N=0}^\infty&
%\frac1{N!} \int \d x_1 \cdots \d x_N
%f(x_1) \cdots f(x_N) 
%\nn\\&\times
%\hat{\psi}^\dagger(x_1) \cdots \hat{\psi}^\dagger(x_N)
%\vac. 
%\end{align}
%In each subspace with fixed number of particles $N$, 
%the coherent state is the product of one-particle state associated with the classical soliton $\fcl(x)$.% 

%%%%%%%%%%%%%%%%%%%%%%%%%%%%%%%%%%%%%%%%%%%%%%%%%%%%%%%%%%%%%%%%%%%%%%%%%%%
%\subsection{Time evolution}
{\it Time evolution. }
Let us proceed to the time evolution of the coherent state. 
According to the principle of quantum mechanics, 
the coherent state $|\fcl\ket$ is time-evolved as
\begin{align}
|\fcl, t\ket:=e^{-\i\hhll t}|\fcl\ket=e^{\ah(-t)}\vac, 
\end{align}
where
\begin{align}
\ah(t):=\int\d x \[\fcl(x) \hpd(x,t)-\fast(x) \hp(x,t)\]. 
\end{align}
The expectation value of the field operator at time $t$ is given by
\begin{align}
\bra\fcl, t|\hp(x)|\fcl, t\ket
=\bra\fcl|\hp(x, t)|\fcl\ket, 
\end{align}
which is {\it not} equal to the classical soliton $\fcl(x, t)$. 
Here let us introduce the ``{\it classically}" time evolved coherent state 
$|\widetilde{f,t}\ket$ as 
\begin{align}
&
|\widetilde{f,t}\ket:=e^{\at(t)}\vac, 
\\&
\at(t):=\int\d x \[\fcl(x,t) \hpd(x)-\fast(x,t) \hp(x)\], 
\end{align}
whose expectation value describes the exact time evolution of the classical soliton 
\begin{align}
&
\bra\widetilde{f,t}|\hp(x)|\widetilde{f,t}\ket=\fcl(x,t).
\end{align}
At the initial time $t=0$, they are equal to each other 
\begin{align}
\ah(0)=\at(0)=\ah, \quad
|\fcl, t=0\ket=|\widetilde{f,t=0}\ket=|\fcl\ket. 
\end{align}
They are evolved in time according to 
\begin{align}
\del_t\ah(t)=\frac1\i[\ah,\hhll], \quad
\del_t\at(t)=\{\ah,E\}. 
\end{align}
As a quantity representing the difference between quantum and classical states, 
we introduce a function $r(t)$ as
\begin{align}
r(t):=\bra\fcl, t|\widetilde{f,t}\ket
=\vvac e^{-\ah(-t)} e^{\at(t)} \vac, 
\end{align}
which starts from 1 and dacays to 0. 
At an infinitesimal time $\Delta t$, we have
\begin{align}
&
\ah(-\Delta t)=\ah-\Delta t \frac1\i[\ah,\hhll]=:\ah-\Delta t F, 
\\&
\at(\Delta t)=\ah+\Delta t \{\ah,E\}=:\ah+\Delta t G. 
\end{align}
Using Baker-Campbell-Hausdorff formula, we can explicitly evaluate $r(t)$ in the form
\begin{align}
r(\Delta t)=1-\i c\Delta t\int|f|^4\d x+c\,\mathcal{O}(\Delta t^2). 
\end{align}
In the case of $c=0$, the overlap $r(t)$ remains 1 for all $t$ and 
the time evolution of the coherent state exactly coincides with the classical soliton. 
The cancellation of the linear terms suggests that 
the instability of quantum soliton originates from the 
nonlinear term in the Hamiltonian \eqref{hhll}. 

%%%%%%%%%%%%%%%%%%%%%%%%%%%%%%%%%%%%%%%%%%%%%%%%%%%%%%%%%%%%%%%%%%%%%%%%%%%
%\subsection{Conclusion}
{\it Conclusion. }
In this letter, we constructed the field theoretical coherent state \eqref{fket} 
and discussed the quantum-classical correspondence 
in the integrable field theory. 
We consider here the case of the one-dimensional Bose gas as an example. 
However, similar arguments are possible in the case of other integrable field theories. 

The expectation values of the field operator with respect to the coherent state 
is identical to the classical soliton at initial time. 
%in which function is included in coherent state. 
However, the unitary time evolution of the coherent states 
breaks the coherent property due to the nonlinearity. 
Consequently the expectation values of the field operators with respect to this state 
do not coincide with the classical solutions anymore. 
We discussed the difference between quantum and classical 
time evolutions of coherent states by evaluating the overlaps between these states. 

%The results, for any theory has no difference of first order of the time evolution. 
%Therefore the coherent state for the classical solution has stable behavior. 
%Moreover, we can calculate of higher order of this difference for the Hamiltonian 
%is given by finite order polynomial for the fields. 
%Therefore we can show that the complete information of the 
%time evolution of the quantum-classical correspondence.

To elucidate the relationship between 
the coherent states and the previously constructed quantum bright and dark solitons 
are the next problem to be studied. 

%%%%%%%%%%%%%%%%%%%%%%%%%%%%%%%%%%%%%%%%%%%%%%%%%%%%%%%%%%%%%%%%%%%%%%%%%%%

\end{document}